\documentclass{report}
\usepackage{setspace}
\usepackage{pdfpages} 
\usepackage{amssymb,graphicx,color}
\usepackage{amsfonts}
\usepackage{latexsym}
\usepackage{a4wide}
\usepackage{amsmath}

\usepackage{geometry}
\usepackage{changepage}
\usepackage{amsmath}

\title{{\vspace{-14em} \includegraphics[scale=0.4]{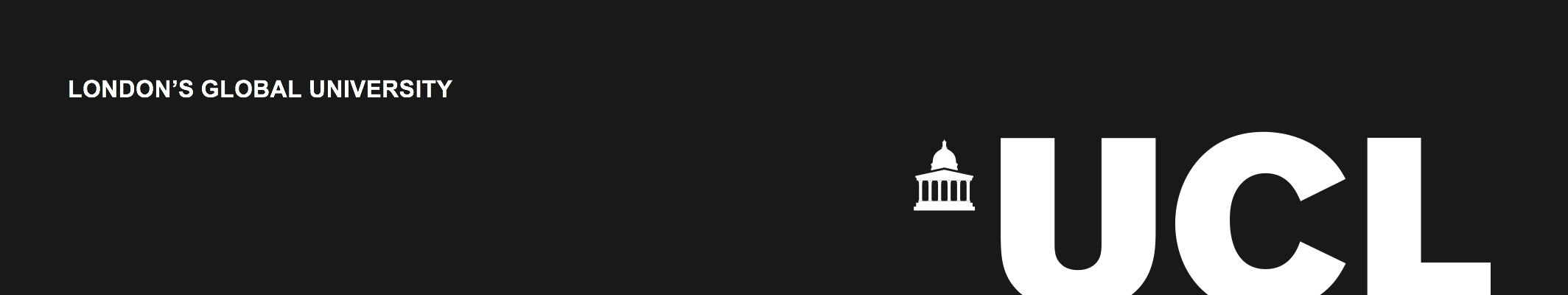}}\\
{{\Huge Generalized Categorisation of Digital Pathology Whole Image Slides using Unsupervised Learning}}\\
}
\date{Submission date: 05 May 2020}
\author{Mostafa Ibrahim\thanks{
{\bf Disclaimer:}
This report is submitted as part requirement for my Computer Science degree at UCL. It is substantially the result of my own work except where explicitly indicated in the text.
\emph{} The report may be freely copied and distributed provided the source is explicitly acknowledged
\newline  
}
\\ \\
BSc Computer Science \\ \\
Supervisor: Kevin Bryson}

\begin{document}

\onehalfspacing
\maketitle
\begin{abstract}

This project aims to break down large pathology images into small tiles and then cluster those tiles into distinct groups without the knowledge of true labels, our analysis shows how difficult certain aspects of clustering tumorous and non-tumorous cells can be and also shows that comparing the results of different unsupervised approaches is not a trivial task. The project also provides a software package to be used by the digital pathology community, that uses some of the approaches developed to perform unsupervised unsupervised tile classification, which could then be easily manually labelled.
\newline

The project uses a mixture of techniques ranging from classical clustering algorithms such as K-Means and Gaussian Mixture Models to more complicated feature extraction techniques such as deep Autoencoders and Multi-loss learning. Throughout the project, we attempt to set a benchmark for evaluation using a few measures such as completeness scores and cluster plots.
\newline

Throughout our results we show that Convolutional Autoencoders manages to slightly outperform the rest of the approaches due to its powerful internal representation learning abilities. Moreover, we show that Gaussian Mixture models produce better results than K-Means on average due to its flexibility in capturing different clusters. We also show the huge difference in the difficulties of classifying different types of pathology textures.

\end{abstract}
\tableofcontents
\setcounter{page}{1}
\chapter{Introduction}

\section{Motivation and Project Overview}
Due to the recent advancement of microscopes, a subfield of pathology has emerged, Digital Pathology.This subfield introduces a realm of possibilities for automation of diagnosis and medical analysis as it digitizes any microscopic slide. Digitized slides are called Whole Image slides and can be viewed under various magnification levels ranging from 20X to 100X. Whole Image slides (WSI) unlike normal images have huge sizes that are usually around 3GB per WSI which makes it challenging to transfer them, analyse them and for example train Deep Neural Networks on them. To overcome the difficulties of WSIs in this project, we used a framework called OpenSlide\cite{openslide} to segment and breakdown a WSI into various small tiles.
\newline 

Histopathology is the study of tissue under microscopic examination to analyse the manifestations of disease.Two very popular stains used in this study are hematoxylin and eosin which help pathologists in distinguishing between normal cells and cancerous nuclei due to specific patterns of condensation of those stains.
\newline

In recent years, there have been a lot of successful applications of machine learning and Artificial Neural Networks (ANNs) in the field of Digital Pathology \cite{deep}, performing various tasks such as classifying cancerous tissues with higher precision than Pathologists and much more.
\newline

This project focuses on using generalised unsupervised techniques to cluster those small tiles into various distinct groups such as cancerous tissue, normal tissue, muscle, adipose, … etc (depending on the presence of those groups in a dataset). The resulting clusters are critical to the success of many different classification algorithms as most of them need data labels for the training process which in the digital pathology field is very hard to acquire, as the labelling process requires pathology expertise and a lot of time. The project does not make any assumptions about the type of cancer of the WSI which increases its value greatly as it can be used on any type of cancer.
Ideally, given a set of tiles, the algorithm would be able to give out a result similar to such figure.
\newline

\hfill\includegraphics[scale=.25]{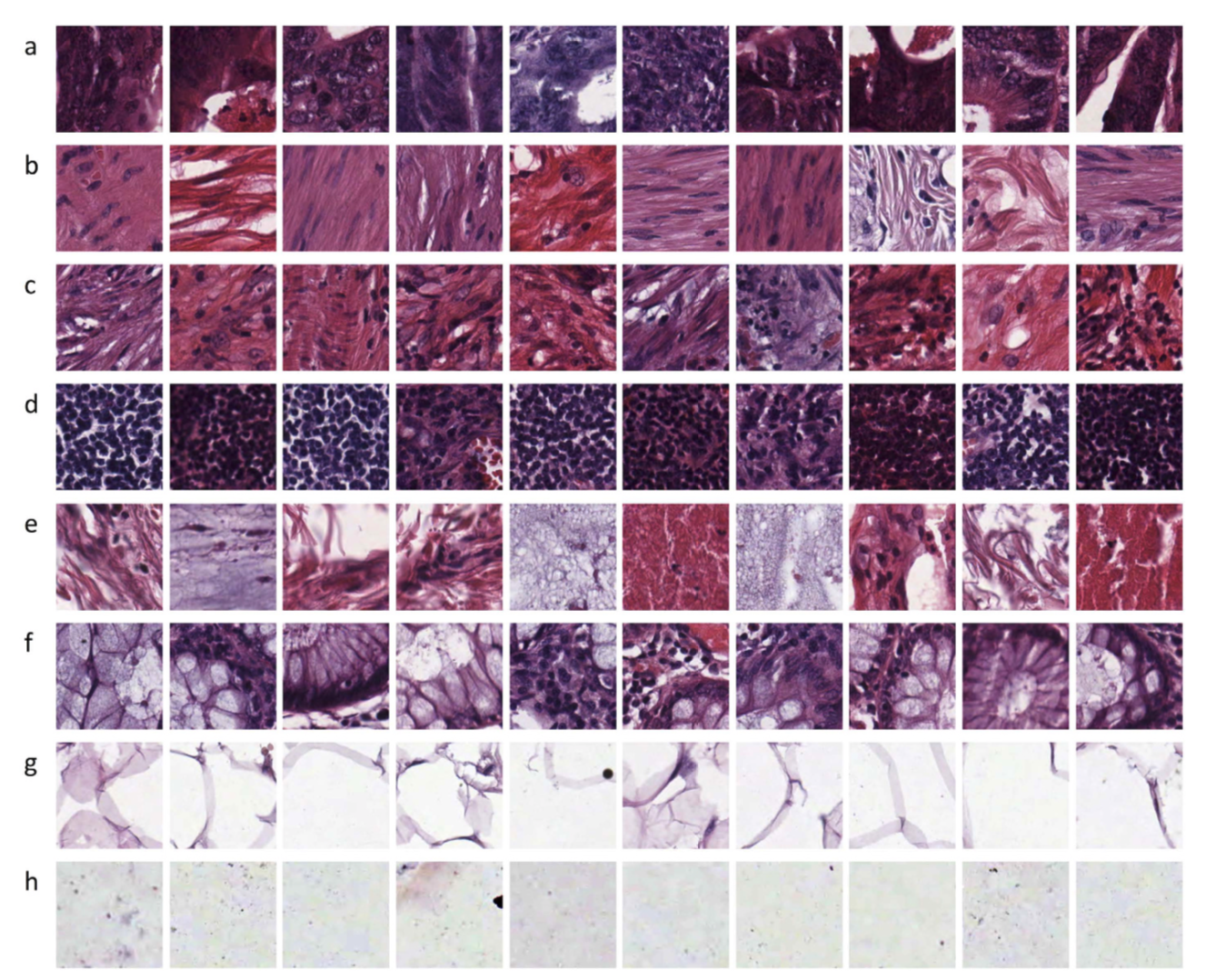}\hspace*{\fill}
\begin{center}
     Figure 1.1\cite{cluster-2}: Tiles showing different tissue types. "(a) tumour epithelium, (b) simple stroma, (c) complex stroma (stroma that contains single tumour cells and/or single immune cells), (d) immune cell conglomerates, (e) debris and mucus,
(f) mucosal glands, (g) adipose tissue, (h) background."\cite{cluster-2}

\end{center}

\section{Project aims and Goals}

The project has the following aims and goals:
\begin{enumerate}
\item Automate most of the process of labelling digital pathology tiles with the proposed approaches.

\item Provide supervised machine learning algorithms with semi-labelled datasets (supervised algorithms need labels to learn feature mappings unlike unsupervised algorithms)

\item Develop a very general algorithm that would work for a wide range of cancers and tissue.

\item Use various distinct approaches ranging from classical machine learning clustering algorithms to advanced deep learning approaches such as Autoencoders and test their performance to achieve the best results.

\item Provide a simple Command Line Interface (Software Package) for users to simply run the algorithm on their WSI dataset. This will allow users to make use of massive (Terabytes) datasets. Most WSIs are processed on server machines or HPC clusters with batch processing or console-only access.

\item Identify various problematic properties of digital pathology images that can make clustering algorithms perform poorly and attempt to get around those properties.

\end{enumerate}

\section{Project approach} 
Throughout the project, two powerful clustering algorithms were used, K-Means, Gaussian Mixture Models. Furthermore, a subfield of unsupervised learning is dimensionality reduction, which aims to reduce the dimensions of the feature space of the data without losing as much relevant information as possible, this helps the machine learning algorithms to perform better and faster \cite{curse}. In this project, 3 dimensionality reduction algorithms will be used, Principal Component Analysis (PCA), t-distributed Stochastic Neighbour Embedding (t-SNE), and Autoencoders.
\newline

The project focuses on using 4 different approaches to solve the proposed problem:
\begin{enumerate}
\item Classical machine learning approach
    \begin{enumerate}
    \item Reducing dimensionality of the full feature space
    \item Clustering on RGB and H\&E values
    \end{enumerate}
\item Deep learning approach
    \begin{enumerate}
    \item Dimensionality reduction with Convolutional Autoencoders
    \item Semi-supervised Convolutional Autoencoders
    \end{enumerate}
\end{enumerate}
All of those approaches differ in the dimensionality reduction aspect, however, they all use the same two machine learning clustering algorithms, this design choice aims to represent the differences of the performance of those algorithms for each approach.
\newline
\newline
The evaluation of each approach and comparing them was one of the main challenges of this project since it is unsupervised (does not use labels). The chosen solution comprises two different measures. The first measure is a visual plot that chooses 10 random images from each cluster and plots them in a similar way to Figure 1.1 to show the similarities between those images in each cluster. The second one provides a concrete measure similar to accuracy called completeness using a dataset that has already been labelled.
\newline
\section{Report structure}
In the next chapter the report discusses several motivations for using the chosen approaches and why they proved to be successful in other similar projects. Then the algorithms used are explained briefly with an emphasis on their loss functions. For each approach that was done, an appropriate relevant analysis was conducted which will be explained thoroughly in order to compare it with the other approaches. Finally, we conclude the discussion with a summary of the work done and entail future work describing several ideas that might be effective to tackle this problem.

\chapter{Background and Literature review}

\section{Brief introduction to digital pathology and Histopathology}
Histopathology is mainly concerned with examination of biological tissues to observe the appearance of diseased cells and tissues on a microscopic levels. Several histopathology techniques have been proven effective when it comes to diagnosing cancer such as applying different types of stains. In section 3.4 we focus on H\&E stains, those stains make the cells’ nucleus purple and the cytoplasm pink, this allows pathologists to more easily identify cell structures and thus diagnose cancer \cite{Leeds}.   
\newline

\hfill\includegraphics[scale=1.7]{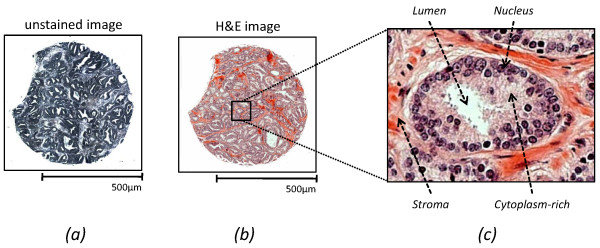}\hspace*{\fill}

\begin{center}
Figure 2.1 \cite{HE}: H\&E Staining highlights important features for diagnosis by improving contrast of tissues.
\end{center}

Cancer is the second leading cause of death world-wide, responsible for around 15\% of deaths\cite{death}. Therefore automating diagnostic processes using machine learning (ML) has become one of the most significant medical tasks. Cancer occurs when the cells start dividing and growing uncontrollably, moreover, it is worthy to note that solid tumors are not simply clones of cancer cells \cite{cancer}, they are extraordinary organs that are made up of several cell types, their texture is not consistent, which makes identifying them a hard problem for both pathologists and ML algorithms.
\newline

More precisely, in this project, we try to address the problem of multi-class texture analysis, in this process we make very few assumptions about the type of cancer and thus the algorithm is generalised. In medical context, the term texture refers to certain properties of the internal structure of images for example rough versus smooth.\cite{texture}
\newline

The main disadvantage of using supervised algorithms to tackle digital pathology problems is that once the algorithm is trained on one type of cancer it only gives good accuracies for that type, however for unsupervised learning, clustering and Dimensionality reduction algorithms have no knowledge of the labels so they are not catered to only one type of cancer.


\section{Dimensionality Reduction and Feature Extraction}
\subsection{Principal Component Analysis}
Most machine learning algorithms are improved greatly in terms of robustness and interpretability through dimensionality reduction (DR) algorithms. PCA has been used greatly to eliminate redundant dimensions that don’t necessarily hold valuable information. It does so by choosing components that have the most variance (the eigenvectors).Typically those components hold all of the valuable information (features) needed for algorithms \cite{DR}. The main theory behind PCA is that the dataset sits on a d dimensional space, where we require only d coordinates to describe a single point on a hyper plane, however the dataset is originally presented in an m dimensional space where m is much greater than d.
\newline

PCA works through Projected Variance Maximisation where the target is to project the original m dimensions into d dimensions such that the sample variance is maximally preserved.This d dimensional subspace encapsulates the directions along which the data varies the most. The loss function is the called the reconstruction error and can be described as
\newline

\hfill\includegraphics[scale=.15]{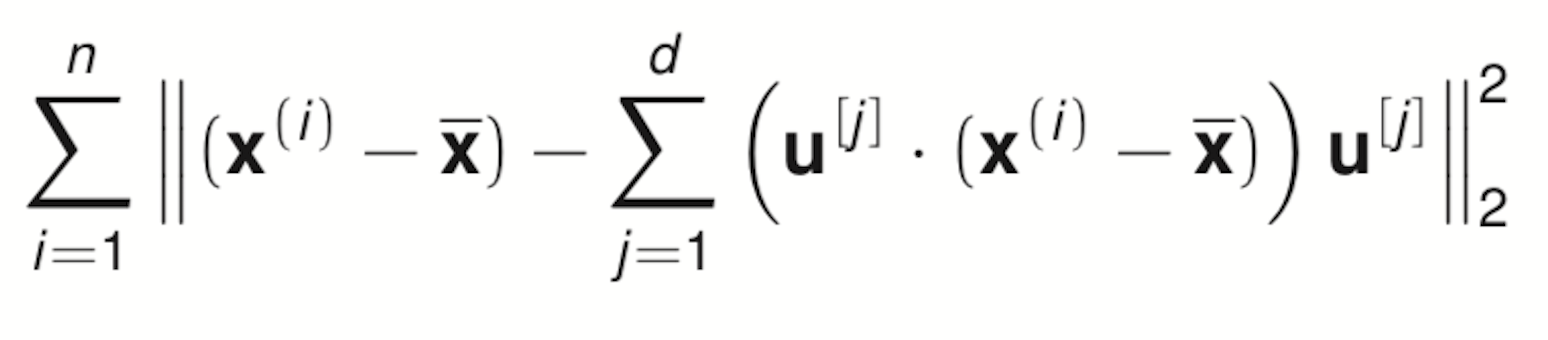}\hspace*{\fill}
\begin{center}
\end{center}

Where $\bar{x}$ is the mean of the project data, u is a set of basis vectors than span the data subspace d and main objective is minimizing the 2 norm squared for points 1 to n (the whole dataset).
\newline

PCA was used by Kamath and Mahato\cite{DR} to classify oral pathology tissues.They observed that using PCA improves the accuracy of algorithms such as k-means nearest neighbors which works in a similar way to unsupervised k-means. PCA has also been used to compress medical images without losing valuable information from those medical images \cite{DR2}.
\newline

\subsection{t-distributed Stochastic Neighbour Embedding}
T-SNE on the other hand is a different type of DR as it is not used in compression but rather visualisation.T-SNE plots have been used significantly in unsupervised projects to see if the features separate the data into natural clusters \cite{sne1}.
\newline

Unlike PCA, t-SNE is a non-linear algorithm that falls under the class of manifold learning / nonlinear dimensionality reduction. The main idea of T-SNE is constructing 2 probability distributions of the dataset, one for the original high dimensional dataset and one for the lower dimensional space (2D) and then as typical machine learning algorithms, it minimises the loss between those 2 distributions through Kullback-Leibler divergence which measures the divergence (difference) between 2 probability distributions and is defined \cite{KL} as 

\hfill\includegraphics[scale=.4]{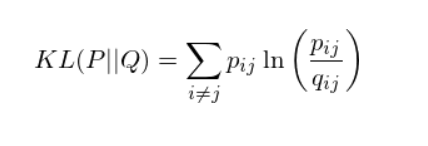}\hspace*{\fill}
\begin{center}
\end{center}

where P is the true distribution of the data and Q is the approximated low dimensional distribution.

\section{K-Means and Gaussian Mixture Models review}

Both K-means and Gaussian Mixture Models (GMMs) are partitional iterative algorithms that use the Expectation Maximization (EM) algorithm, which essentially performs maximum likelihood estimation (maximizing likelihood function of a distribution) in the presence of latent variables (inferred variables). The expectation step is where the algorithm estimates the parameters for the latent variables, then the maximization step optimises those estimated parameters.Those 2 steps are repeated until convergence.
\newline

This method of optimization has proven to be very effective, for instance in this paper \cite{EM}, they were using an EM based algorithm to "decompose histological images into independent tissue classes (e.g. nuclei, epithelium, stroma, lumen) and align the color distributions for each class independently" which provides significant motivation for using K-Means and GMMs here. Furthermore, GMMs were used in this paper \cite{GMM} to perform mitosis detection in breast cancer histological images, they have found out that GMMs helped in reducing false positives and had greater interpretability compared to other more sophisticated machine learning algorithms. 
\newline

A variation of K-Means was used by Takayasu et al.\cite{K-Means} to attempt replacing a supervised task with an unsupervised one which is similar to the scope of this project. They were trying to do unsupervised pathology image segmentation, they first use a variation of K-Means to learn internal representations of pathology images. Then, they used the conventional K-Means for clustering those representations, they also manage to achieve a Normalised Mutual Information score of 0.626.This measure falls under the category of evaluation metrics for unsupervised algorithms using labels which we will be discussed later.

K-Means uses the Euclidean distance as the similarity measure for data points. It starts off with randomly initialising k centroids, where k is a hyperparameter, after that it attempts to make the inter-cluster data points as similar as possible (minimize the distances) while also keeping the clusters as different as possible. To do this, it uses the following loss function
\newline

\hfill\includegraphics[scale=.3]{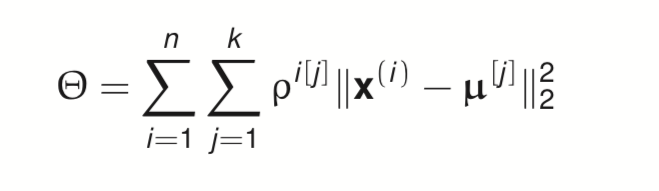}\hspace*{\fill}
\begin{center}
\end{center}

Where $\mu$\textsuperscript{[j]} is cluster number j, $\rho$\textsuperscript{i}\textsuperscript{[j]} is a boolean indicator variable used to indicate whether the data point i belongs to cluster j. The target of K-Means is simply to minimise the sum of squares of the distances of each instance to its closest centroid, to optimise this loss function. The Expectation Maximisation algorithm is used (which will also be used in Gaussian Mixture Models). This algorithm is guaranteed to converge but not necessarily to a global optimum. 
\newline


As with regards to GMMs, Gaussian mixtures are essentially a mixture of k features (hyperparameter) where each k is assumed to have a Gaussian distribution. The main target of this algorithm is to estimate the parameters that best fit the k features combined. Those parameters are the mean u that defines the centre, the covariance matrix (matrix of pairwise variances) defining the width and the mixing probability that defines how big or small the Gaussian is. 
\newline




While K-Means uses  the concept of hard clustering where each point either belongs to a cluster or not, GMMs uses soft clustering, where each point has some probability that it belongs to a cluster. Moreover, K-means is only able to achieve a good clustering result for cases where the data points form spherical clusters, one the main reasons of using GMMs here is the fact that they can learn ellipsoidal clusters too, which we will show later that they sometimes outperform K-Means.

\section{Deep learning}

Deep learning is based on Artificial Neural Networks (ANNs) \cite{DL} which are biologically inspired models that are used to approximate functions (especially non-linear ones). Those networks are composed of several hidden layers (the more layers the deeper the network), one input layer (in this project, it's the pixels of an image) and one output layer. Each layer is composed of several neurons, those neuron connections are dependent on the architecture of the network. The main process by which ANNs work is called learning and this relies on 3 main functions, the loss function, the activation function and the optimization algorithm. 
\newline

The activation function is a function applied at every neuron. The simplest activation function is the linear function where the output is the weighted sum of the input vector plus the biases. These functions have many types saturated; squeezes the input to a certain range, such as Sigmoid [0,1], non-saturated; doesn't squeeze, such as ReLU. There are also linear and nonlinear activation functions. 
\newline

The loss function defines the measure by which network is performing, so typically the loss would be large at the beginning and if the learning process of the ANN is working as expected it would reduce this loss after every epoch (learning iteration). The choice of the loss function differs greatly according to the learning task at hand, however the one that will be discussed later in the Autoencoders section is the Mean Squared Error (MSE) function.
\newline

The optimization algorithm is used to reduce the loss function during the training process, typical optimization algorithms rely on the concept of gradient descent where after every epoch the ANN would be taking a closer step towards a minimum that minimizes the loss function. A variation of the gradient descent used here is called Adam (Adaptive moment estimation) \cite{Adam}.
\newline 

Gradient descent (GD) \cite{GD} is essentially an iterative process that relies on the concept of partial derivatives (derivative with respect to one variable in a multivariable function). GD updates the loss function’s weights on every iteration using the parameters from the last iteration. The learning rate determines the size of the learning steps it takes and the gradient of the loss function.
\newline

In this project, we will focus on a type of neural network called Autoencoders (AE). They have been greatly used to encode images, reducing their dimensionality greatly and recording there most essential features in bottlenecks. AE architectures typically look like this:
\newline

\hfill\includegraphics[scale=.22]{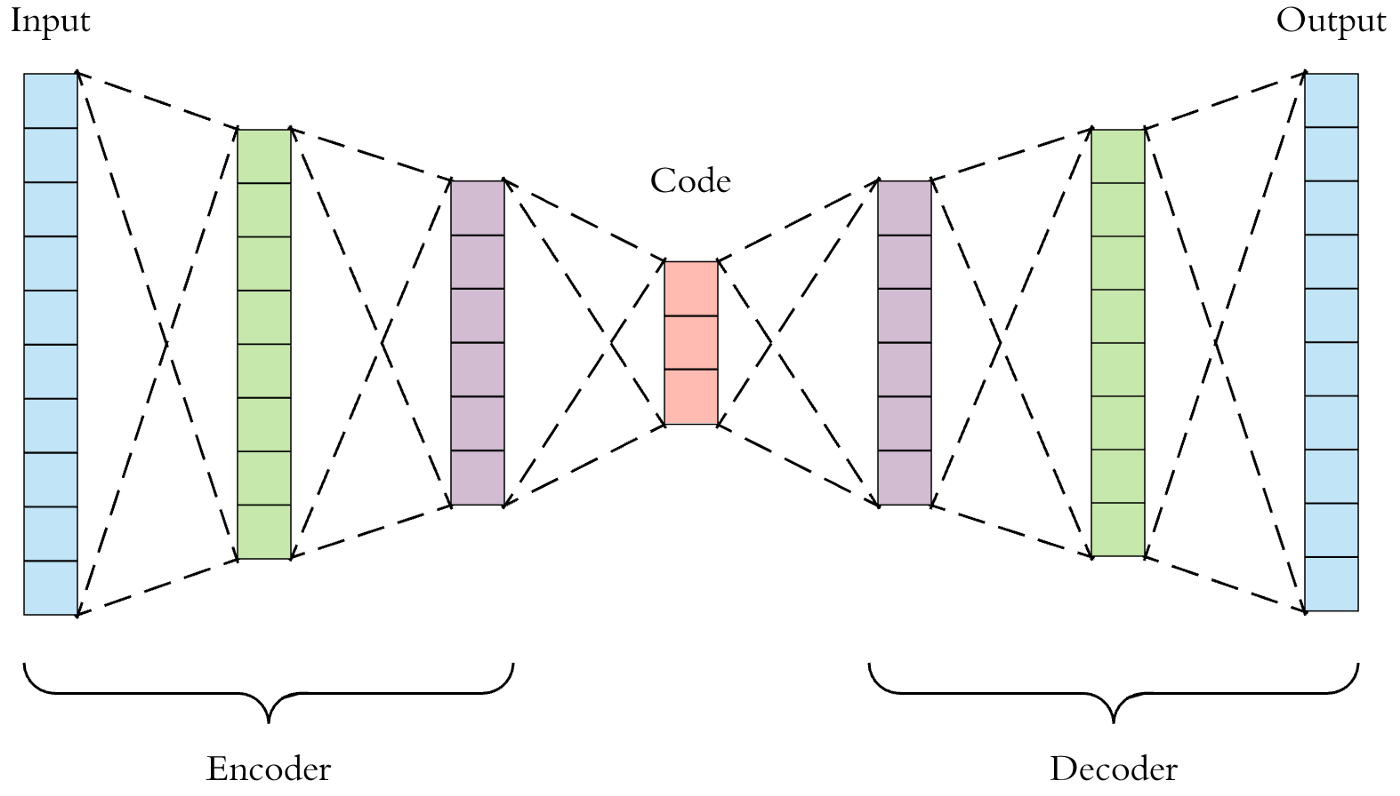}\hspace*{\fill}
\begin{center}
Figure 2.2\cite{aevis}: Autoencoder Architecture Visualisation
\end{center}

The main motivation for the encoding-decoding process is that if the decoder is able to reconstruct the image from the internal code with minimal error, then this indicates that only this internal representation (code) is necessary to represent the image, i.e it represents the most significant features of the image. This can be seen as a feature extraction method.
\newline

Autoencoders have been applied in digital pathology in many papers.XU J and Xiang L\cite{ae} used multiple Autoencoders (stacked) to learn high-level features from pixel intensities to identify distinguishing features of nuclei. Also, they were used by Ruqayya Awan and Nasir Rajpoot\cite{ae2} to overcome stain variations, capturing features for deriving feature maps of histology WSIs and applying powerful representation learning.

\section{The multi-class texture problem}

Jakob Nikolas Kather and Cleo-Aron Weis \cite{cluster-2} address exactly the same problem we are addressing here, except that they use supervised algorithms such as radial basis Support Vector Machines. Only very few studies have addressed the multi-class problem and for specific types of cancer there are very few published results. The multi-class texture problem refers to the problem of clustering multiple tissue types given a tissue sample (in this case 8). Moreover, the fact that pathologists can disagree on the labelling on tissues \cite{pathologists} \cite{pathologists2} means that the labelled dataset isn't 100\% accurate and thus accuracy measures such as completeness would be slightly compromised.
\newline

Figure 1 helps us visualise the difficulty of the multi-class texture problem especially in an unsupervised setting, since some rows have highly varying structures and shapes as rows a, c and e. Throughout this paper, they have applied their experiments twice, once for only 2 classes (tumour/stroma), and once for 8 classes (the multi-class problem). This process shows the true difficulty of the 8 classes problem such that they were able to achieve around 97\% accuracy for  the 2 classes and only 87\% for the 8 classes problem. We will show a similar analysis in the fourth approach where the completeness score for 2 classes is much higher than that for 8 classes.
\newline

In this paper, instead of using DR methods which apply automatic feature extraction, they use image processing techniques to extract features then apply supervised algorithms on them. For instance, texture descriptors, local binary patterns (visual descriptors), Gabor filters which analyze whether there are any specific frequency content in the image.
\newline


\chapter{Methods and Results}

\section{Datasets}

There were 3 main datasets used in this project:
    
\begin{enumerate}
    \item Kaggle\cite{Kaggle} dataset
    \newline
    The tiles in this dataset represent Histopathologic scans of lymph node sections. Furthermore, each tile has binary labels, either tumorous or non-tumurous, these labels were only used in training the semi-supervised approach. This dataset is based of the PatchCamelyon \cite{pcam} dataset, one of the most used datasets in histopathology. We decided to use the version found at Kaggle as they presented a version of the PatchCamelyon dataset which didn’t have duplicates and those can affect the training of neural networks.
    
    \item University Medical Center Mannheim (UMCM) Dataset
    \newline
    This is the dataset from the paper \cite{cluster-2} that was reviewed in section 2.4, 
    it is much smaller in terms of size than Kaggle's one because the number of classes here is 8 rather than 2,this was valuable for evaluation, benchmarking and for the semi-supervised approach. This dataset was originally obtained from the pathology archive at the University Medical Center Mannheim (Heidelberg  University, Mannheim, Germany).
    
    \item TCGA \cite{tcga} Dataset (WSIs)
    \newline
    This dataset was only used for testing the whole software package, especially the tiling process. It mainly consisted of 5 WSIs of colon cancer.
\end{enumerate}

Table 3.1 summarizes the sizes of the datasets, both Kaggle's dataset and UMCM's dataset were split as 80\% for training and 20\% for testing.
\newline

\begin{center}
\begin{tabular}{ |c|c|c|c|c| } 
 \hline
 Dataset & Training tiles & Testing tiles & Total tiles & Classes \\ 
 \hline
 Kaggle & 220,025 & 57,458 & 277,483 & 2 \\ 
 \hline
 UMCM & 4,200 & 800 & 5,000 & 8 \\ 
 \hline

\end{tabular}
\end{center}

\begin{center}
\begin{adjustwidth}{2.6cm}{}
\begin{tabular}{ |c|c| } 
 \hline
 TCGA & 10 WSIs for software package testing \\ 
 \hline

\end{tabular}
\end{adjustwidth}
\end{center}

\begin{center}
Table 3.1: Datasets Breakdown
\end{center}

\section{Evaluation metrics}

Completeness Scores and Cluster Plots are the main 2 measures that we used for comparing and evaluating different approaches. Completeness score is intended to be more of a quantitative measure, while the cluster plots are just intended to be a visualisation aid of the performance of the clustering and DR algorithms.
\newline

\textbf{Completeness score}:  This score is not the same as accuracy, since the unsupervised algorithms have no knowledge of the ground truth labels; they are simply just placing every group of data points in a cluster, but they don’t know what this cluster is.
\newline

A clustering result satisfies completeness if all the data points that are members of a given class are elements of the same clusters, this metric is independent of the absolute value of labels \cite{sk}
\newline

More formally, completeness is \cite{completeness}
\newline

\hfill\includegraphics[scale=.35]{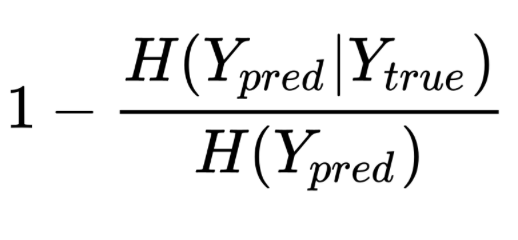}\hspace*{\fill}
\begin{center}
\end{center}

Where H is the entropy function \cite{entropy}, Y\textsubscript{pred} is the predicted label, and Y\textsubscript{true} is the true label, this score stems from mathematical combinatorics. This metric ranges from 0 to 1, where 1 means that all members of a given class are assigned to the same cluster.
\newline

\textbf{Cluster plots}:  To better visualise the results, we developed a plotting system that samples 9 tiles randomly (to avoid any bias) from each cluster and plots them on one row where each row resembles a cluster. This was developed before we got access to the labelled dataset. This visualisation was intended to be similar to Figure 1.
\newline

In Figure 1.1, there were 8 rows, the first seven represent different tissue types and the last one is just background (of the microscopic slides). 

\begin{itemize}
    \item The first 3 rows (a,b,c) representing tumour, simple stroma and complex stroma are the most challenging structures to be learned.
    \item the last 2 rows (g,h) (adipose and background) are the easiest due to their simple structure and their consistency
    \item rows 4-6 (d-f) have medium difficulty in comparison to other rows, also note that row e contains 2 classes debris and mucus.
\end{itemize}

For every approach we will show the corresponding cluster plot with a similar analysis pertaining to the consistency of rows and the performance compared to the difficulty of the class/row.

\section{Working with the Full Dimensional Space}

Each image in UMCM’s dataset was 150x150 RGB (3 color channels), this gives us a total number of dimensions of 67,500. Our initial thought was that clustering algorithms perform poorly in extremely high dimensional spaces \cite{curse}, so we had to find a way  to reduce those dimensions somehow. Principal component analysis has proven to be very  effective as discussed in section 2.2.1 and so we decided to use it. The second step was to cluster on the output of PCA which is intended to have the most varied features of the image.
\newline

The dimensionality of the reduced dimensional space of PCA had to be changed and tested such that we can get the maximal reduction in dimensions with the minimal loss in variance, after several trials shown in Figure 3.1, we found that we can preserve 99\% of the variance and get a 97\% reduction in the number of dimensions.
\newline

\hfill\includegraphics[scale=.4]{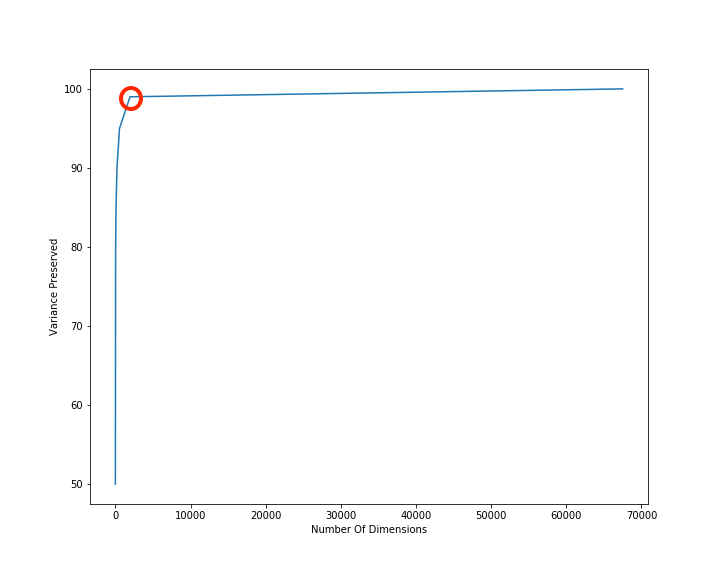}\hspace*{\fill}
\begin{center}
Figure 3.1 shows the variance preserved for different number of dimensions of the reduced dimensional space. The Full Dimensional Space is 67,500. The red circle shows the optimal point of reduction (which is around 2,000 dimensions).
\end{center}

To find the best hyperparameters that fit the Gaussian Mixture Models and K-means, we used Scikit-learn's \cite{sk} GridsearchCV which essentially tries all of the hyperparameters from a given selection of hyperparameters, evaluates them against the completeness score and use K-Fold cross validation to evaluate the training and validation errors on different folds for the data, then gives out the best hyperparameter setting.
\newline

Since we already know that our dataset has 8 different types, we set the n\_clusters to 8, however for other hyperparameters such as the covariance type for Gaussian Mixture models, initialisation algorithm for K-means, and random initialisation state for both of them, we used gridsearchCV.
\newline

\subsection{Approach results}

After performing Gridsearch for this approach, K-means managed to outperform GMMs, the best K-Means had a completeness score of
0.50 on the training set and 0.54 on the testing set.
\newline

Figure 3.2 shows the resulting cluster plot for this approach: 
\newline

\hfill\includegraphics[scale=.3]{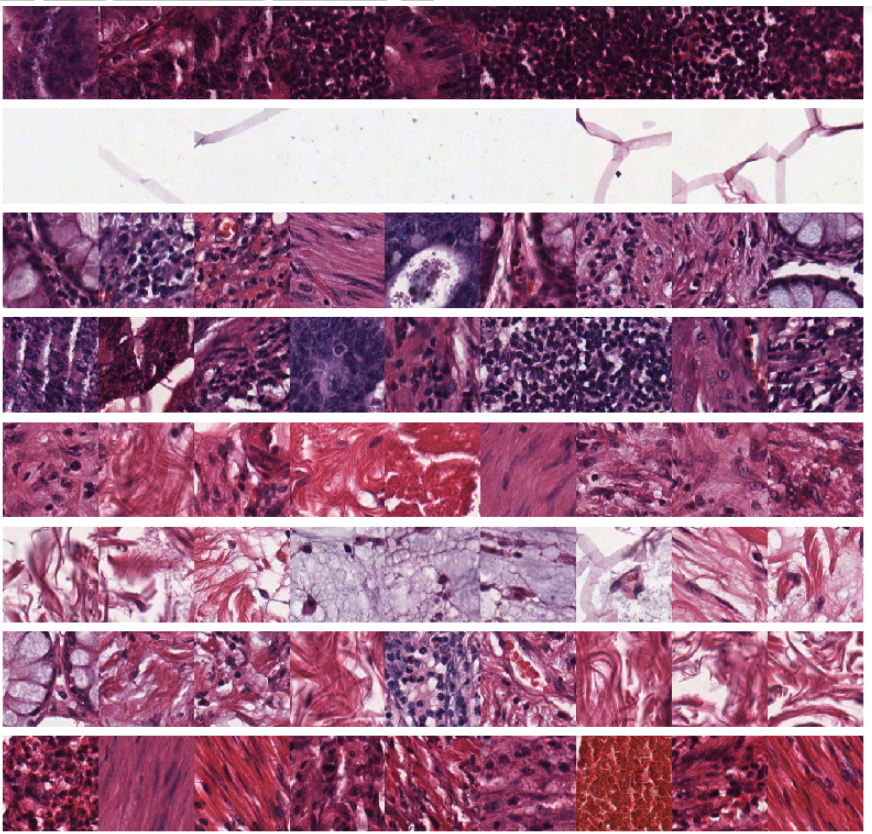}\hspace*{\fill}
\begin{center}
Figure 3.2: Cluster plot of the Full Dimensional space approach.
\end{center}

From the plot we can see that there is good definition on the first 2 rows, however it did not manage to entirely separate the adipose tissues from the background ones for example.
\newline

It seems that the reduced dimensional space is not providing the clustering algorithms with features that give good cluster separation (as the rows are fairly mixed). Therefore we had to either find a better way of doing automatic feature selection or manually select the features which brings us to the next approach at which we extract the Hematoxylin \&  Eosin, and the mean RGB values from images and use those as features to cluster on.

\section{Extracting RGB and H\&E stains as features}

In this approach we loop through each tile, and record its average RGB and H\&E values. This means that for each tile we only have 5 features (so 5 dimensions). The H\&E extraction is done using scikitlearn’s \cite{sk} "rgb2hed" function which converts the color space from RGB to H\&E.
We also observed that after using PCA to reduce the dimensions from 5 to 2, the clustering algorithms were able to separate the background cluster (which is entirely white as in Figure 1.1 from the adipose cluster).
\newline

Gaussian Mixtures here managed to outperform K-Means with a completeness score of 0.63 for the testing set and 0.59 for the training set, the best fitting covariance for GMMs is a spherical one.
\newline

\hfill\includegraphics[scale=.5]{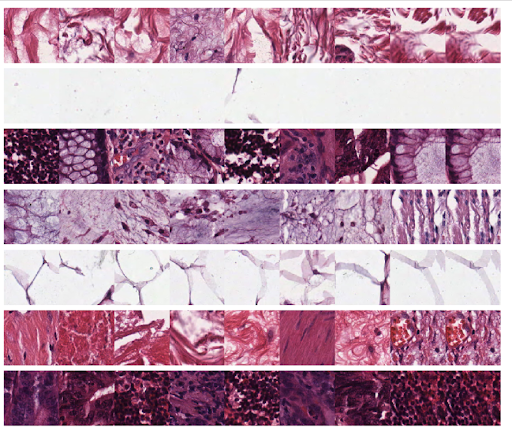}\hspace*{\fill}
\begin{center}
Figure 3.3: The cluster plot for the RGB and H\&E approach.
\end{center}

The plot show in Figure 3.3 is definitely an improvement over the previous approach, we can see that row 2 is clearly background, row 5 is adipose, row 1 is mostly stroma. Other rows have a few outliers. Since after the dataset is now 2D (after PCA), we can visualise it through Figure 3.4:
\newline

\hfill\includegraphics[scale=.25]{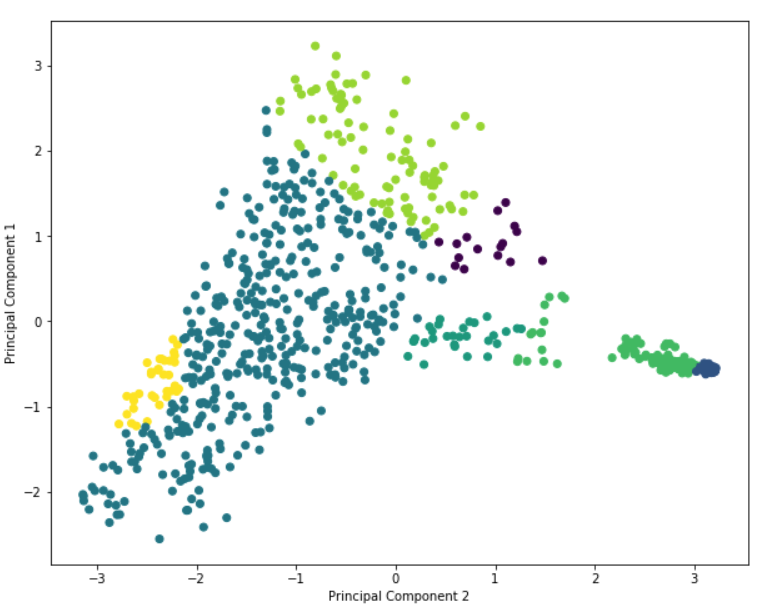}\hspace*{\fill}
\begin{center}
Figure 3.4: RGB and H\&E approach Principal Components Plot. This shows the separation of the clusters given that those two principal components were used as features.
\end{center}

Although the separation is not well defined for every cluster, we can see that the 4 clusters on the right have a fairly good separation.



\section{Deep Feature Extraction}

In this section, we start using a more sophisticated approach which is deep learning to extract features and reduce dimensions. We start off by reviewing Convolutional networks which lay down the  foundation for Convolutional Autoencoders that we used. Then we discuss the approach details and results. After that, we move to the final approach which is multi-loss semi-supervised Convolutional Autoencoders in which we attempt to improve the performance of the Autoencoders with the use of labels.
\newline

\subsection{Review: Convolutional Networks}

\hfill\includegraphics[scale=.64]{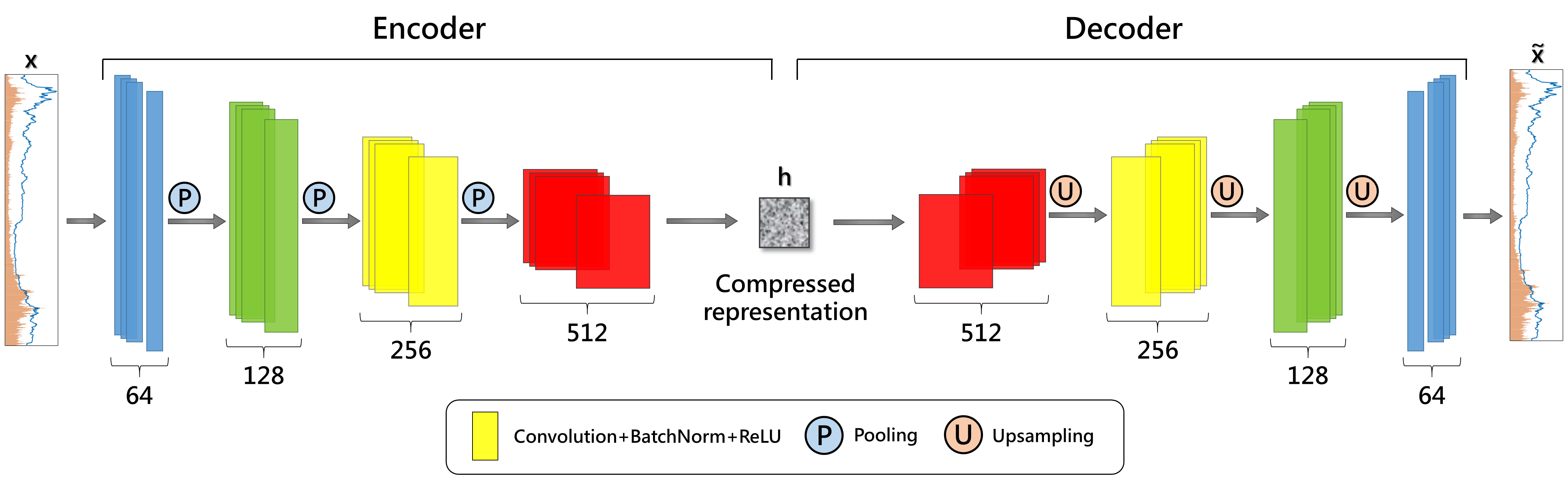}\hspace*{\fill}
\begin{center}
Figure 3.5\cite{fg}: Convolutional Autoencoder Example, in an unsupervised setting, the compressed representation which essentially contains the most significant features of an image is used in clustering.
\end{center}

Convolutional Autoencoders use the same concepts found in Convolutional Neural Networks (CNNs), these concepts are essentially techniques to extract as many relevant features from images as possible. The three foundational techniques to achieve this feature extraction task are convolutions, pooling and non-linear activations. Stacking those three layers and repeating them several times allows CNNs to learn powerful low-level features at the beginning and higher-level features as the network gets deeper. 
\newline

A convolution in terms of image processing, is an operation where a kernel slides over the image pixels, performing element wise multiplication with the part of the input, and then summing up the results into a single output pixel. Varying the values of the kernel changes the features that are being extracted from the image, such as shapes, edges and textures. In deep learning, neural networks manage to learn the values of the kernel that minimizes the loss function, which in turn leads to extracting the optimal features from images, those optimal features are called feature maps and they result from the repeated application of the kernel on the image that is being convolved.
\newline

Another significant aspect of Convolutional networks is pooling layers, these layers provide an effective approach to “down-sampling” the images and summarizing the features in a feature map. In our approaches, we used max pooling which returns the maximum value from the portion of the image covered by the kernel. Pooling provides a dimensionality reduction aspect to our network as it reduces the size of each feature map by a certain factor (which in our network is 2).
\newline

Since convolutions are linear (dot product), CNNs make use of non-linear activation functions in order to learn more complicated non-linear kernel weights. In our networks (approaches 3 and 4), we used Leaky-ReLU \cite{ReLU} as our activation function, given by this equation:
\begin{center}
    $y_i = \begin{cases}
                        \frac{x_i}{a_i} & x_i < 0 \\
                        x_i & x_i \geq  0
                      \end{cases} $ \\
    where $\alpha_i$ is a hyperparameter 
\end{center}

Non-linear activation functions have significantly improved CNNs \cite{CNN} making them state of the art for image recognition tasks. There is a huge body of evidence motivating the use of CNNs in digital pathology, with so many papers using CNNs to perform several digital pathology tasks.

\begin{itemize}
    \item First, it was used by Gertych and Swiderska-Chadaj \cite{CNN1} to accurately distinguish distinct growth patterns of lung adenocarcinoma (a type of lung cancer) from  WSIs. The CNN managed to achieve an overall accuracy of 89\% for distinguishing 5 tissue classes.
    
    \item They were also used by Pegah Khosravi and Ehsan Kazemi \cite{CNN2} to distinguish between lung cancer, bladder cancer and breast cancer with accuracy of approximately 92\%, discriminating between different tissue types is a similar task to the one we are trying to accomplish here. There results clearly demonstrate the power of CNNs in learning pathology-relevant feature maps from WSIs that make the classification with this accuracy possible.
    
    \item Lastly, in this paper \cite{CNN3} they used a variation of CNNs to perform digital pathology segmentation and improve tumor detection performance on challenging lymph node metastases dataset, which is the same dataset that we are using to train the networks (the PatchCamelyon/Kaggle dataset).

\end{itemize}

\textbf{Autoencoder loss functions}: In this project, we experimented with 2 different loss functions that were used to evaluate the reconstruction of the autoencoder the first one is the MSE loss function which measures the squared difference in pixel values of the reconstructed image and the original image, and outputs the average difference it is given by this equation \cite{MSE}:
\begin{center}
    \[L(x, y) = \frac{1}{N} \sum_{i=0}^{N}(x_i - y_i)^2\]
    where x is the original image and y is the reconstructed image, N is the number of training tiles
\end{center}

The second loss function that we experimented with is Structural Similarity Index Measure
(SSIM)\cite{SSIM}, which is more involved than MSE in terms of complexity, we are not going to discuss how SSIM works as it is not entirely relevant and for the sake of the length of this report. However, SSIM is better than MSE in capturing the structures and shapes found in images, since it does use the values of pixels in the same way as MSE, it is not as sensitive to noise as MSE \cite{SSIM2} ,it is given by this equation:
\begin{center}
    \[SSIM(x,y) = \frac{(2\mu_x\mu_y + c_1)(2\sigma_{xy} + c_2)}{(\mu^2_x + \mu^2_y + c_1)(\sigma^2_x + \sigma^2_y +c_2)}\]
    where $\mu$ is average, $\sigma$ is variance ($\sigma_{xy}$ is covariance), c is added to stabilize the division.
\end{center}

\subsection{Approach details}

All of the autoencoders were trained on a local workstation with 1 NVIDIA GTX 1070 GPU using Keras (an open-source machine learning library developed by Google).In this approach the network was trained on Kaggle's dataset, but since the size of this dataset was too large compared to the depth of our network, we did not need to use all of it to get good MSE reconstruction results. We used 49,000 tiles for training and 15,000 for testing. Also,we moved through several Autoencoder architectures which will not be discussed in detail, but essentially the 2 final Autoencoder had the following configuration:


\begin{center}
\begin{tabular}{ |c|c|c| } 
 \hline
 Loss & bottleneck size & Number of Parameters being learned \\ 
 \hline
 MSE & 1,152 & 365,283 \\ 
 \hline
 SSIM & 1,296 & 382,739  \\ 
 \hline

\end{tabular}
\end{center}

\begin{center}
    Table 3.2 Autoencoders
\end{center}

We used early stopping to adjust the number of epochs, which is a mechanism that stops the Neural Network from training once the derivative of the loss function stops changing for 10 epochs, this also helps the Neural Network to not overfit as it does not continue on learning the noise from the images. For further robustness, we monitored training and testing error graphs along with training and testing reconstruction errors that will be shown later in order to avoid overfitting or underfitting. We also used a few Batch Normalisation layers which effectively normalises the values of activations in a hidden layer, this improves the training speed greatly, and reduces overfitting due to regularization effects.
\newline

We can visualise reconstruction error as shown in Figure 3.6, these are just 10 random images sampled for the testing set:
\newline

\hfill\includegraphics[scale=.25]{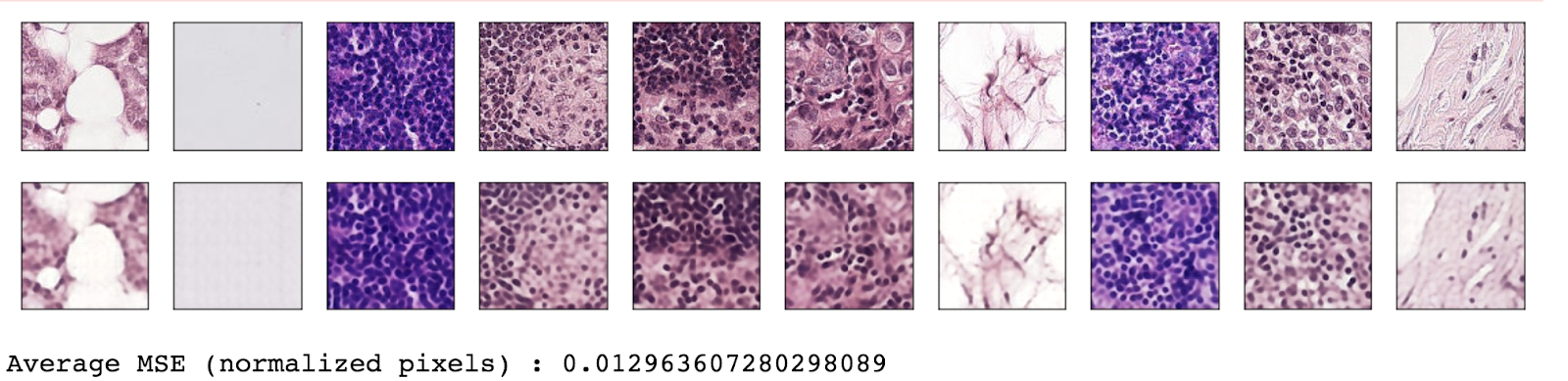}\hspace*{\fill}
\begin{center}
Figure 3.6: Autoencoder reconstruction and MSE, first row shows the original images, and second row shows the reconstructed images after the encoding-decoding process. This Autoencoder was trained using MSE as the loss function.
\end{center}


To evaluate the autoencoders, we calculated the average training and testing completeness scores. The results are shown in this Table 3.3:

\begin{center}
\begin{tabular}{ |c|c|c| } 
 \hline
 Loss & Training Completeness & Testing completeness \\ 
 \hline
 MSE & 0.63 & 0.65\\ 
 \hline
 SSIM & 0.61 &  0.65\\ 
 \hline

\end{tabular}
\end{center}

\begin{center}
 Table 3.3 shows the average MSE completeness scores for the 2 autoencoder variations, the first one uses MSE as the loss function and the second one uses SSIM as shown in the previous table. The completeness scores shown were measured by GMMs for both autoencoders (since they were higher than K-Means).
\end{center}

Since the MSE autoencoder had better completeness scores, we will only show its clustering results. After encoding the images using the autoencoder, we used PCA with 98\% variance preserved to further drive down the number of dimensions. After that we attempted to fit K-Means and GMMs using these new compressed features, and same as with the previous approaches, we used GridsearchCV to find the hyperparameters that best fit the training data and minimize the validation error through cross-validation. 

The cluster plot can be seen in Figure 3.7:
\newline

\hfill\includegraphics[scale=.6]{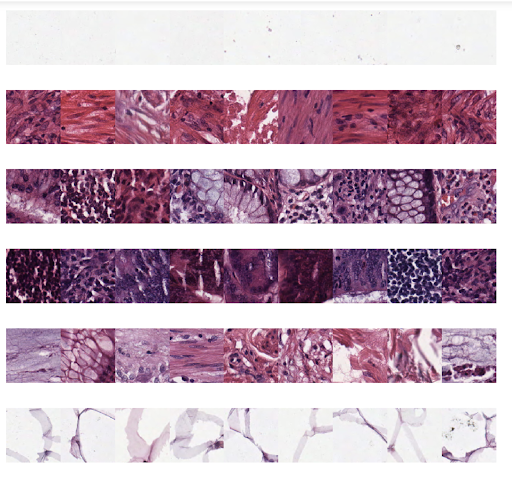}\hspace*{\fill}
\begin{center}
Figure 3.7: Autoencoder Cluster Plot. Shows the result of reducing dimensions using the convolutional autoencoders, then PCA, then clustering the reduced feature space using Spherical Gaussian Mixture Models.
\end{center}

We can see that there are only 6 rows, although the clustering algorithms were fitted with 8 clusters, they only managed to find 6. However, this produces the best completeness score and as we can see from the plot, the adipose tissue, background and simple stromas (rows 1,2 and 6) were clustered correctly. Row 4 seems to be a mixture of tumours and conglomerates.
\newline

Furthermore, we used T-SNE as a mechanism to visualise the data space to see if we can observe clusters as mentioned in section 2.2.2. t-SNE has several hyperparameters that required constant changing, but the most important one is perplexity which measures the effective number of neighbours.Figure 3.8 shows the T-SNE plot for different perplexities:

\hfill\includegraphics[scale=.5]{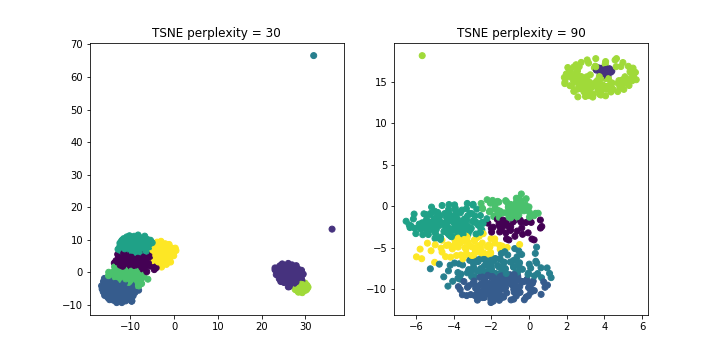}\hspace*{\fill}
\begin{center}
Figure 3.8: T-SNE plot, we can see clear separation for at least 2 different types of clusters. However, we still need better features that will aid in better separation, which brings us to our final approach, semi-supervised autoencoders.
\end{center}

\subsection{Review Semi-supervised Mutli-loss learning and Classification}

At this stage we started looking into more novel and different approaches to tackle the problem at hand.One way to do this was to add a new feature to our autoencoders which is a second bottleneck that attempts to do classification/regression using the data labels while the first bottleneck is using MSE/SSIM to minimize the reconstruction loss.This is know as Multi-task learning (MTL) \cite{MTL} where the Neural Network attempts to learn 2 tasks simultaneously. Any change to the weights and biases in the shared layers is guided by the current loss in all outputs in such environment. 
\newline

MTL has been used widely in many fields including digital pathology, for instance it was used in this paper \cite{MTL2} to perform classification and grading of histological images. One of the reasons why MTL usually improves performance is that it induces a form of regularization by requiring the Neural Network to perform well on both tasks.It also aids in improving generalization by forcing the shared layers to find commonalities among distinct tasks. In our case this represents the first bottleneck learning an internal representation with features that helped in performing the classification in the second bottleneck, which are more valuable and relevant features. This can be seen as a semi-supervised learning approach such that we are using labels to learn the features, however we are not using the labels to predict the class for a tile (which is a supervised task).
\newline

\hfill\includegraphics[scale=.24]{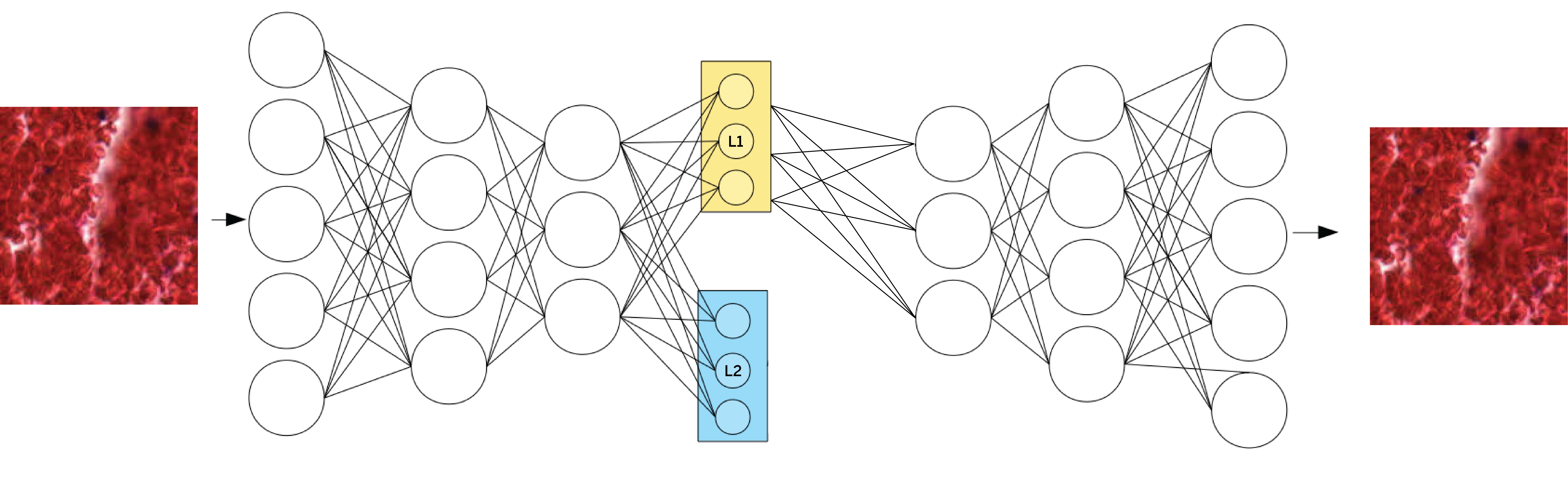}\hspace*{\fill}
\begin{center}
Figure 3.9: Semi-Supervised Autoencoder, L1 is the reconstruction loss, and L2 is the classification loss
\end{center}

This approach required testing with different datasets and different settings, we first present a network that uses MSE both as reconstruction and regression loss functions, this network is trained on Kaggle's dataset. Then we present 2 networks that both use categorical cross entropy as a classification loss function \cite{cce}, one of them uses MSE to measure reconstruction and the other one uses SSIM (as with Convolutional Autoencoders), and thus these 2 networks were trained on UMCM's dataset. The reason for training the networks on 2 different datasets is to provide results that prove the difficulty of the multi-class texture problem mentioned in section 2.5.
\newline

As for the networks trained on UMCM's dataset, we had to use a technique known as Data Augmentation to "increase" the size of the dataset. Data Augmentation performs several operations during network training to provide more images for training, for instance, we performed three 90 degrees rotations on every image. The reason for using this technique on UMCM's dataset is that there were only 4200 training tiles (compared to 49,000 for Kaggle's one). Data Augmentation has proven effective in digital pathology \cite{DA1} in cases with limited datasets. It increases the dataset's size which prevents networks overfitting and increases the overall performance of the network. "It increases the robustness of the models against input variability without reducing the effective capacity and may also enable learning more biologically plausible features"\cite{DA2}.
\newline

\textbf{Activation and loss functions used}: 

\begin{itemize}
    \item Here, we used softmax as an activation function for the final fully connected layer. Softmax acts as a sigmoid function in the sense that it squashes the output in the range 0 to 1, but it also ensures that the sum of outputs along channels (as per specified dimension) is 1.Therefore for each output there are K (where K is the number of classes) values, each values represents a probability that this data point belongs to this class.It is given by this equation \cite{softmax}:

    \begin{center}
        \[\sigma(y)_i = \frac{exp(y_i)}{\sum_{j=0}^{K}exp(y_j)} \]
        where exp is the exponential function, and i = 1,...,K
    \end{center}
    
    \item For 2 classes (Kaggle dataset) we used MSE to do regression.
    
    \item For 8 classes (UMCM's dataset) we used Categorical Cross entropy, which is one of the most widely used loss functions to perform multi-class classification \cite{cce}, which is given by this equation
    
    \begin{center}
    \[L(x, y) = - \sum_{j=0}^{N}\sum_{i=0}^{K}x_{i_j} * log(y_{i_j})\]
         where y is the predicted value, N is the number of training samples
    \end{center}

\end{itemize}

The purpose of using the labels here in this way is to not actually use them when predicting the labels of tiles since this is an unsupervised project and prediction using labels is supervised. The purpose is to enhance the features learned by the autoencoder (through learning 2 tasks).

\subsection{Testing and Evaluation}

To make comparison between the approaches more clear, we decided to perform the same clustering and PCA procedure here as the one performed in the convolutional autoencoders section. The only addition is that to further evaluate the new autoencoders, we now evaluate their supervised prediction accuracies (in addition to the reconstruction errors). This helps us indicate whether they are actually learning useful features about the images or not. The metric used for this task is known as the "f1-score", which is given by this equation \cite{f1-score}:

\begin{center}
        \[2 * \frac{precision * recall}{precision + recall} \]
\end{center}

This table presents a summary of the architectures, their results and the reconstruction (Rec) errors:

\begin{center}
\begin{tabular}{ |c|c|c|c|c|c|c|c| } 
 \hline
 Loss & bottleneck size & Parameters & Classes & Training f1 & Testing f1 & Training Rec & Testing Rec \\ 
 \hline
 MSE & 2,592 & 726,283 & 8 & 0.56 & 0.55 & 0.019 & 0.019 \\ 
 \hline
 SSIM & 2,592 & 814,251 & 8 & 0.45 & 0.44 & 0.016 & 0.016 \\ 
 \hline
 MSE & 576 & 466,964 & 2 & 0.92 & 0.92 & 0.014 & 0.014 \\ 
 \hline

\end{tabular}
\end{center}

 Table 3.4 shows a fairly interesting result, which is that the f1-score for 2 classes is much higher than the f1-score for 8 classes, this proves the difficulty of the multi-class texture problem. It is also worth noting that the multi-class problem will be harder than the 2-class because it is easier to choose the correct class when only 2 classes are involved (50\% random chance) compared to when 8 classes are involved (12.5\% random chance of getting it correct).
\newline 

Since the cluster plot for this approach did not show anything particulary new, we decided to only show the T-SNE plot:

\hfill\includegraphics[scale=.5]{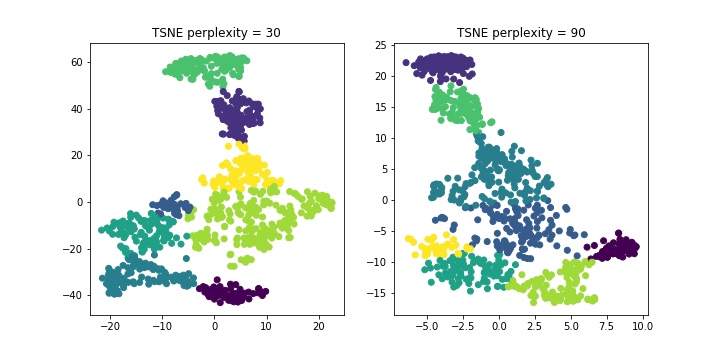}\hspace*{\fill}
\begin{center}
Figure 3.10: T-SNE plot, shows fairly better separation for the clusters, especially for the one with perplexity 30, this is a good indication that the features learned are relevant and aid in clusters separation
\end{center}

\section{Summary}

The most quantitative way to compare and evaluate the results of the many approaches that we have tried, is to show their completeness scores, as shown here:

\hfill\includegraphics[scale=.5]{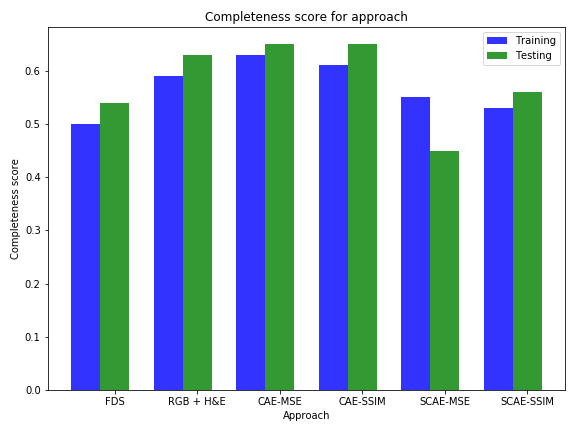}\hspace*{\fill}
\begin{center}
Figure 3.11: Results Summary: 1) FDS is the Full Dimensional Space approach in section 3.3. 2) RGB + H\&E is the second approach, discussed in section 3.4.
\newline
3\&4 are the convolutional autoencoders (CAE) with the 2 different loss measures. 5\&6 are the semi-supervised convolutional autoencoders (SCAE) with the 2 different loss measures.
\end{center}

We can see that the Convolutional Autoencoder using MSE has the best training and testing completeness score. We expected that the semi-supervised autoencoders would have the highest completeness scores, however, it seems that to achieve higher f1-scores (and thus improving the learned features), the networks need to have more complicated and deeper architectures. Although our networks depth and complexity increased to up to around 815,000 parameters they were still unable to fully capture the essence of the features. To further investigate this, we show the f1-score for each of the classes for the Semi-supervised MSE autoencoder in Figure 3.11:

\hfill\includegraphics[scale=.35]{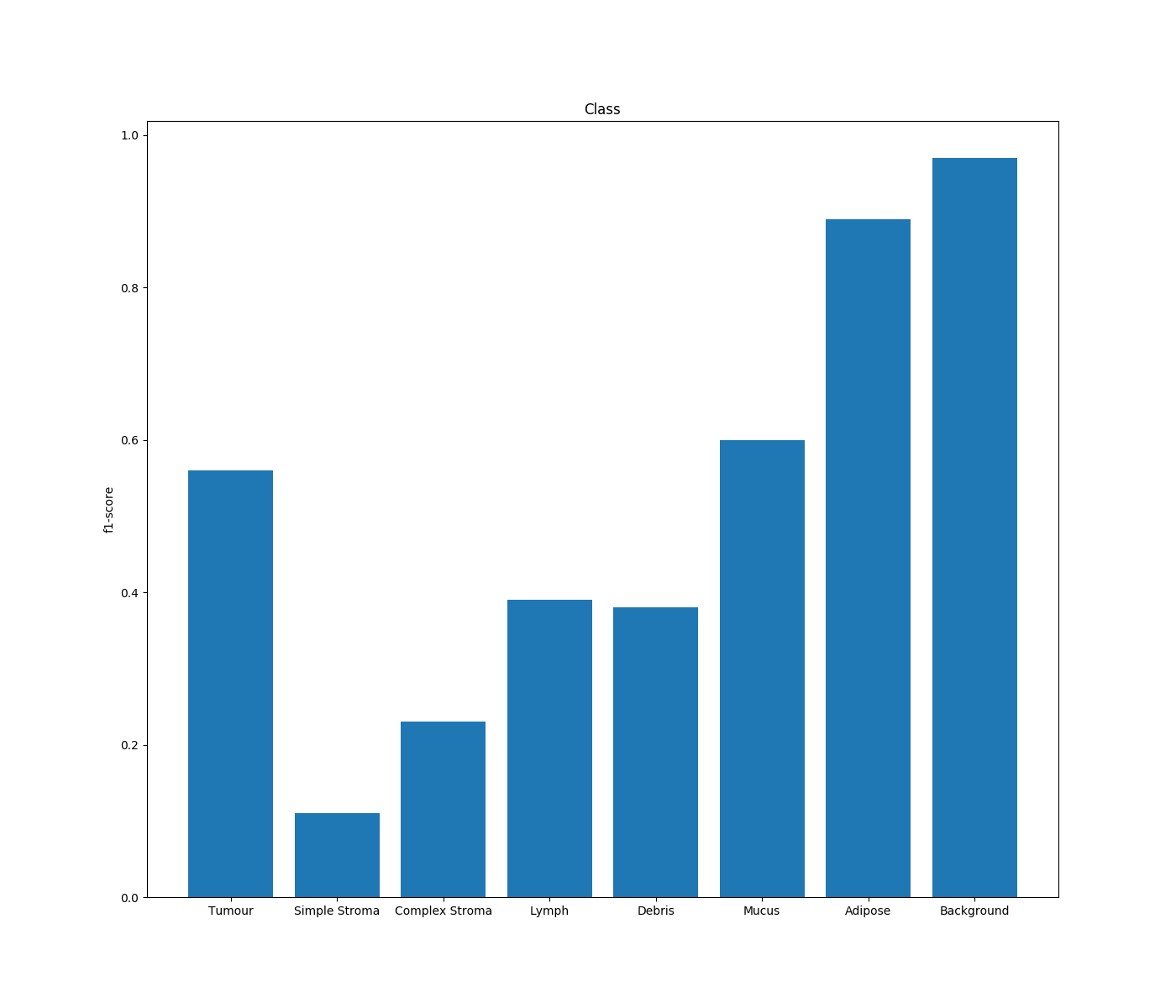}\hspace*{\fill}
\begin{center}
Figure 3.12: The plot shows that the f1-score for the simple and complex Stroma is very low compared to other classes.
\end{center}

This shows that the autoencoder was underperforming in both of the Stroma classes and therefore, the network needs to be deeper. The most probable reason for this is the complexity of the shapes of stromas, which are supportive tissues of tumours, and also they seem to have a wide variability in their shapes which prevented our networks from learning their features. This is fairly shown within most of the cluster plots, where the stroma row is usually mixed with tumors or immune cell conglomerates.
\newline

While deep networks offer certain advantages in terms of modelling complex functions and outperforming learning tasks compared to classical machine learning algorithms, we can see that a simple approach such as the RGB $+$ H\&E also achieved relatively a high completeness score. We can also see that the difference between the SSIM and MSE autoencoders is not very large although SSIM offered significantly lower construction errors as shown in table 3.3
\newline

It is also worth mentioning that PCA improved the completeness scores of all of the approaches, as we used it after the autoencoders (and before clustering) by around 15\%. It also seemed to aid the separation of classes, particularly the adipose and background classes. Moreover, for all of the approaches except FDS and SCAR-MSE, gaussian mixture models seem to achieve higher compeleteness scores than K-Means, this is probably due to their flexibility in capturing the non-spherical nature of the clusters, whereas K-Means is only good at capturing spherical clusters.

\section{Software Package}
To make this code easily usable for the digital pathology community we wrapped it in a simple software package. The main use of this package is to breakdown a Whole Image Slide and separate its tiles into groups. This is implemented as sub-directories (where each subdirectory is a cluster). Since it is not practical to include all of the approaches developed in the software package we picked the CAE-MSE approach which had the best completeness as option 2 and the RGB$+$H\&E approach as option 1. The user is free to select which option they prefers. To test this software package, we tried several WSIs from the TCGA dataset as discussed in section 3.1.

\subsection{OpenSlide}
Openslide has become recently popular in the word of digital pathology, since Whole slide images are huge (with dimensions over 50,000 x 50,000 pixels), they are typically very hard to learn with Convolutional Neural Networks, and since they can not be resized significantly without losing valuable information due to the accuracy and precision needed to make a classification, the images are split into tiles to be processed separately and if needed later stitched together.
\newline

Openslide provides a comprehensive API that deals with several issues in the tiling processes such as splitting the tiles and losing contextual information \cite{openslide}. It is a C library that performs the tiling process of the Whole-image slides where it goes through the WSI with a “window” framework dissecting into into small tiles, those tiles are then using in clustering. Moreover, it supports multiple WSI formats such as .svs, .tiff, .svsslide, and others.
\newline

The default Openslide script was slightly modified by Sophie Peacock \cite{Sophie} which mainly provides several small feature upgrades on the current Openslide distribution. Due to time restrictions we decided to use her script. After breaking down a WSI using this script, several subdirectories are created with different magnification levels ranging from 1.25 to 20.0 where each subdirectory has the tiles corressponding to its magnification level.

\subsection{Command Line Interface}

This is an overview of the software package:

\begin{enumerate}
    \item It first uses Openslide to break down the WSI into non-overlapping tiles, then it navigates to the magnification level specified by the used, then if the user selects:
    \begin{enumerate}
    \item Option 1: It extracts the mean RGB and H\&E from each tile and uses a Gaussian Mixture Model to fit the data with n\_clusters specified (estimated by the user).
    
    \item Option 2: It applies 2 sets of dimensionality reduction algorithms before clustering, the first set is the encoding process of an Autoencoder network, the second set further reduces the dimensions by  applying the PCA algorithm, after this feature extraction process has finished, it uses a Gaussian Mixture Model to fit the data.
    \newline
    \end{enumerate}
\end{enumerate}

Then it creates n\_cluster sub-directories in the specified magnification level directory where each sub-directory represents a distinct cluster.It then saves a symbolic link of each of the images corresponding to that cluster in that sub-directory. The symbolic links where a clever design choice, we did not want to copy the images because that would take more time and space. Also, the symbolic links are easily previewable.
\newline

\hfill\includegraphics[scale=.27]{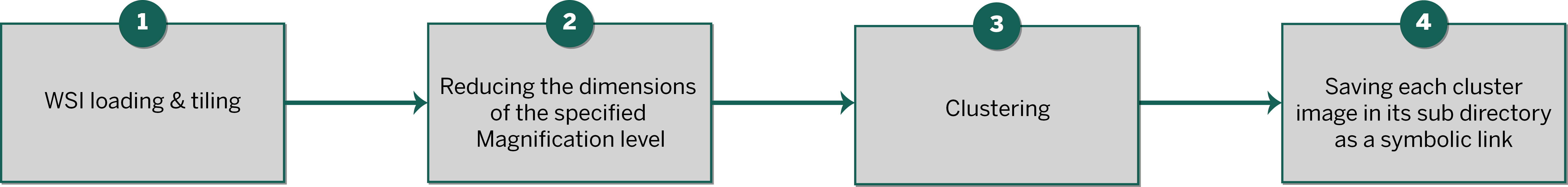}\hspace*{\fill}
\begin{center}
Figure 3.13: System Diagram
\end{center}

The software package is essentially a Command Line Interface (CLI) argument parser, it gives users the flexibility to:

\begin{itemize}
    \item Estimate the number of groups that might be potentially found in their dataset
    \item Choose the magnification level that they want to cluster on
    \item Choose which of the 2 approaches they think might be best
\end{itemize}



\chapter{Conclusion}

\section{Summary of work}
The project shows a wide variety of approaches to tackle the problem at hand. We keep the 2 clustering algorithms consistent throughout the project, but we keep changing the feature selection / elimination techniques. We first start by simply using PCA to apply dimensionality reduction on the Full Dimensional Space. After that we attempt to manually pick out the best features that can represent our dataset which are the RGB and H\&E values. Furthermore, we move on to more advanced dimensionality reduction methods, deep autoencoders, which prove to be our best result. After that, we attempt to boost their performance using labels, to achieve that we use multi-task learning as a mechanism for learning features that are relevant to classify the tiles into their respective classes. Finally, we measure the completeness score and evaluate cluster plots throughout all of our approaches, along with extra measures such as reconstruction errors and prediction accuracies for (semi-supervised) Autoencoders. We keep classic machine learning principles in mind such as checking training and testing errors (overfitting and underfitting) throughout the project. Finally, we apply a few optimisation techniques that are relevant to each approach such as Gridsearch through hyperparameters, batch normalisation, and early stopping.

\section{Critical evaluation of the project's results}

One of the most challenging aspects of a digital pathology project is interpretation of results, pathology tiles have extremely high complexities and similarities. We have attempted to demonstrate the results in various distinct measures both qualitative and quantitative such as cluster plots, completeness scores, prediction accuracies, and reconstruction errors. However, it is still difficult to tell whether this approach outperforms other approaches precisely, especially in terms of generalisation as well, as this project was intended to work for all different types of cancer. We discuss a possible solution to this problem in the Future Work section. We were also using an unsupervised metric known as the "Silhouette Coefficient" to assess the clustering algorithms, also we attempted using a third clustering algorithm known as "Agglomerative Hierarchical Clustering" in this project, but we excluded both of them from the report for the sake of clarity, being concise and the fact that they did not produce interesting results.
\newline

Most of the digital pathology papers use supervised techniques to tackle digital pathology problems. In this project we investigated using unsupervised learning to tackle the multi-class texture problem which is quite a novel technique, with not enough papers using similar approaches. However, we still manage to produce an unsupervised software package that can be used in any supervised digital pathology project to perform unsupervised pre-training to produce distinct clusters. These clusters can then manually be labelled, this saves a lot of time as instead of manually labelling each tiles, the user can just label the whole cluster.
\newline

\section{Future Work}

There are a lot of ideas that we had for this project, but due to time restrictions especially with the Corona Virus situation, we had to halt work on some of them.

\begin{enumerate}
    \item The first idea was to use Variational Autoencoders (VAEs), VAEs work in a similar way to our multi-task learning approach in the sense that they have 2 different bottlenecks, one is the variance of the dataset and one is the mean of the dataset, and then the Autoencoder attempts to perform statistical sampling using those 2 bottlenecks. We are not going to explain thoroughly how VAEs work, but the main outline is that its internal representation offers various properties that aid in clustering and separating out the tiles from each other. We have attempted to develop VAEs, however the solution was under-developed due to time restrictions so we decided not to include the results in this report.
    
    \item The second solution was regarding improving the complexity of the semi-supervised autoencoders so that they can have higher f1-scores for the Stroma classes. A possible way to do this is to add skip connections to the classification bottleneck. Skip connections are extra connections between nodes in different layers of a neural network. They help in traversing the information throughout the network and have been very successful in certain networks known as "Residual Neural Networks".
    
    \item The third idea was regarding non-deep dimensionality reduction algorithms. Some of the T-SNE plots that we observed throughout this project seem to indicate the presence of a complex manifold that dataset lies in. So it might be worth using different manifold learning algorithms or distinct non-linear dimensionality reduction algorithms such as Locally Linear Embedding that can reduce the dimensions of the dataset without losing valuable features.
    
    \item The final idea and one of the most crucial ones is to boost the interpretability of the autoencoders. There is a fair amount of research in visualising activations of Neural networks, and typically medical projects require extensive interpretability figures to prove that the algorithms fully work. An excellent suggestion that was used in a  few digital pathology papers, is a framework known as "DeepDream", this framework aids in visualising activations in CNNs, highlighting image features that led to, for example, the classification of an image. Using this framework on our autoencoders would show the most significant features that they were able to capture and thus improve robustness. Other methods also exist such as Class Saliency maps.
\end{enumerate}

\appendix


\chapter{Code listing and System Manual}
https://github.com/mostafaibrahim17/PathologyFinal
\newline
\newline

The CLI takes several arguments:

\begin{enumerate}
    \item Datapath
    \item Options: Integer option: 1 or 2
    \begin{enumerate}
        \item Option 1: Chooses the manually selected features approach for clustering
        \item Option 2: Chooses the convolutional autoencoder approach, then clusters
    \end{enumerate}
    \item Number of Clusters: Expected number of data types (Estimated by the user)
    \item Magnification Level: The openslide package tiles the WSI into several magnification levels such as 1.25, 2.5, 5.0, 10.0 and 20.0, the user is given an option to choose any of those magnification levels
\end{enumerate}

\end{document}